\documentclass[%
 aip,
 amsmath,amssymb,
preprint,%
]{revtex4-1}

\usepackage{graphicx}% Include figure files
\usepackage{dcolumn}% Align table columns on decimal point
\usepackage{bm}% bold math
\usepackage[utf8]{inputenc}
\usepackage[T1]{fontenc}
\usepackage{mathptmx}
\usepackage{etoolbox}
\usepackage{xcolor}

\draft % marks overfull lines with a black rule on the right

\graphicspath{{figures/}}

\newcommand{\dso}{{D\lowercase{y}S\lowercase{c}O$_3$}}
\newcommand{\aOrtho}{{$[\overline{1}10]_o$}}
\newcommand{\bOrtho}{{$[001]_o$}}
\newcommand{\cOrtho}{{$[110]_o$}}
\newcommand{\etal}{{\it et al.}}

\newcommand{\pto}{{P\lowercase{b}T\lowercase{i}O$_3$}}

\newcommand{\sro}{{S\lowercase{r}R\lowercase{u}O$_3$}}
\newcommand{\sto}{{S\lowercase{r}T\lowercase{i}O$_3$}}

\begin{document}

\title{Nanoscale domain engineering in \sro\ thin films}

\author{Céline Lichtensteiger}
\email[]{Celine.Lichtensteiger@unige.ch}
\affiliation{Department of Quantum Matter Physics, University of Geneva, 24 Quai Ernest-Ansermet, CH-1211 Geneva 4, Switzerland}

\author{Chia-Ping Su}
\affiliation{Université Paris-Saclay, CNRS, Laboratoire de Physique des Solides, Orsay 91405, France}

\author{Iaroslav Gaponenko}
\affiliation{Department of Quantum Matter Physics, University of Geneva, 24 Quai Ernest-Ansermet, CH-1211 Geneva 4, Switzerland}

\author{Marios Hadjimichael}
\affiliation{Department of Quantum Matter Physics, University of Geneva, 24 Quai Ernest-Ansermet, CH-1211 Geneva 4, Switzerland}

\author{Ludovica Tovaglieri}
\affiliation{Department of Quantum Matter Physics, University of Geneva, 24 Quai Ernest-Ansermet, CH-1211 Geneva 4, Switzerland}

\author{Patrycja Paruch}
\affiliation{Department of Quantum Matter Physics, University of Geneva, 24 Quai Ernest-Ansermet, CH-1211 Geneva 4, Switzerland}

\author{Alexandre Gloter}
\affiliation{Université Paris-Saclay, CNRS, Laboratoire de Physique des Solides, Orsay 91405, France}

\author{Jean-Marc Triscone}
\affiliation{Department of Quantum Matter Physics, University of Geneva, 24 Quai Ernest-Ansermet, CH-1211 Geneva 4, Switzerland}

\date{\today}

\maketitle

{\it Abstract - We investigate nanoscale domain engineering via epitaxial coupling in a set of \sro /\-\pto /\sro\ heterostructures epitaxially grown on (110)$_o$-oriented \dso\ substrates. The \sro\ layer thickness is kept at 55 unit cells, whereas the \pto\ layer is grown to thicknesses of 23, 45 and 90 unit cells. Through a combination of atomic force microscopy, x-ray diffraction and high resolution scanning transmission electron microscopy studies, we find that above a certain critical thickness of the ferroelectric layer, the large structural distortions associated with the ferroelastic domains propagate through the top \sro\ layer, locally modifying the orientation of the orthorhombic \sro\ and creating a modulated structure that extends beyond the ferroelectric layer boundaries.}

\section{Introduction}

Ferroelectric polarisation can be used to affect the properties of other materials. This is well know in ferroelectric field-effect transistors for example, where the polarisation surface charge of the ferroelectric film is used to reversibly dope the adjacent layer, as demonstrated in epitaxial oxide thin film heterostructures~\cite{Ahn-Science-1995}.

In this work, a \pto\ layer is sandwiched between two \sro\ layers. In bulk, \sro\ is a ferromagnetic metallic transition-metal oxide and is often used as an electrode in the ferroelectric oxides community~\cite{Eom-Science-1992}. It is also an itinerant ferromagnet with a Curie temperature $T_C$ = 160 K~\cite{Cao-PRB-1997}. In thin films of this material, the formation of complex spin textures can be induced by the ferroelectric polarisation in an adjacent ferroelectric layer. These include ferroelectric proximity effect near the BaTiO$_3$/\sro\ interface giving rise to an emergent Dzyaloshinskii–Moriya interaction, thereby creating robust magnetic skyrmions~\cite{Wang-NatMat-2018}. Most recently, in \sro /\pto\ heterostructures a ferroelectrically induced magnetic spin crystal was observed~\cite{Seddon-NatCom-2021}. 

\sro\ is not only affected by the polarisation in adjacent layers, but also by the epitaxial strain imposed by the substrate~\cite{Koster-RevModPhys-2012}. When grown on \sto , epitaxial \sro\ layers are organized into structural domains, according to six possible orientations of the orthorhombic unit cell with respect to the cubic substrate. The orientation of the orthorhombic unit cell and resulting domains is affected by the steps and terraces at the surface of the \sto\ substrate~\cite{Jiang-MSEB-1998}. The growth of \sro\ onto vicinal planes of miscut \sto\ substrates leads to the privileged development of a majority single domain orientation in which small domains with different orientations are embedded~\cite{Jiang-APL-1998}. Additionally, control of the \sro\ can be achieved not only through the choice of substrate~\cite{Vailionis-APL-2008}, but also by modifying the growth temperature~\cite{Zakharov-JMT-1999}. Further structural domain engineering has been conducted through control of substrate miscut direction, demonstrating a one-to-one correspondence between structural domains and magnetic domains~\cite{Wang-NPJQuantMats-2020}. 

A structural coupling can also be achieved by strain propagation between the different layers themselves. As was shown in \pto /\sto /\pto\ heterostructures on GdScO$_3$ where the structural coupling between the \pto\ and \sto\ layers resulted in periodic polar waves in the \sto ~ \cite{Tang-NanoLetters-2021}, in \sro /\pto\ superlattices  the large local deformations of the ferroelectric lattice are accommodated by periodic lattice modulations of the metallic \sro\ layers with very large curvatures~\cite{Hadjimichael-NatMat-2021}.

At the core of the heterostructure studied here is \pto , a tetragonal ferroelectric with a polarisation developing along the $c$-axis mostly due to ionic displacements. In \pto\ thin films, the orientation of the polarisation and arrangement into domain structures have been theoretically studied~\cite{Pertsev-PRL-1998,Pertsev-PRL-2000,Koukhar-PRB-2001,Li-APL-2001,Jiang-PRB-2014,Chapman-PCCP-2017}, and are described by phase diagrams with regions of different domain configurations as a function of epitaxial strain and temperature (see review by Schlom \etal ~\cite{Schlom-AnnuRevMaterRes-2007}). The domain pattern is also affected by the film thickness~\cite{Kittel-PR-1946} and electrostatic boundary conditions~\cite{Lichtensteiger-NanoLett-2014,Lichtensteiger-NJP-2016}. Complex polarisation configurations in \pto\ have been reported recently in \pto/\-\sto\ superlattices~\cite{Yadav2016,Stoica2019,Hadjimichael-PhD-2019,Hadjimichael-NatMat-2021,Goncalves2019,Das2019,Wang-NatureMaterials-2020}, with simultaneous control of these configurations using electric fields and light, giving rise to novel phenomena like negative capacitance~\cite{Iniguez-NatureReviewsMaterials-2019}. When grown on \dso , \pto\ takes the $a/c$ phase, where the polarisation forms ordered ferroelastic domains ($a/c$ twins), resulting in distortions of the film surface visible by atomic force microscopy~\cite{Vlooswijk-APL-2007,Catalan-NatMat-2011,Nesterov-APL-2013,Lichtensteiger-APLMaterials-2023}.

Whether through ferroelectric polarisation or strain effects, controlling the structure and morphology of the \sro\ thin films is of importance as it will affect the film electronic resistivity via structural and electronic coupling. Here, we study the structural coupling between oxide thin film layers on a set of \sro /\pto /\-\sro\ heterostructures epitaxially grown on (110)$_o$-oriented \dso\ substrates. We establish the direct role that the ferroelastic domain structure in \pto\ plays in the determination of the orthorhombic domain structure in \sro.

\section{Results}

A series of samples was grown by off-axis RF magnetron sputtering on (110)$_o$-oriented \dso\ substrates, with bottom and top \sro\ electrodes of 55 unit cells (u.c.) and \pto\ film thickness of 23, 45 and 90 u.c. (see Section.~\ref{section:Growth} for details of samples growth).

The substrate, \dso , is orthorhombic with room temperature lattice parameters (in $Pbnm$ space group) $a_o$ = 5.443(2) \AA , $b_o$ = 5.717(2) \AA\ and $c_o$ = 7.901(2) \AA ~\cite{Velickov-ZKristallogr-2007}. It is often useful to refer also to the pseudocubic unit cell, where the lattice parameters can be calculated as $a_{pc}$ = $c_{pc}$ = $\frac{\sqrt{a_o^2+b_o^2}}{2}$ = 3.947 \AA , $b_{pc}$ = $c_o/2$ = 3.951 \AA , $\alpha_{pc}$=$\gamma_{pc}$=90$^\circ$, $\beta_{pc}$=2 $\cdot$ $\arctan$(${a_o/b_o}$)=87.187$^\circ$ at room temperature. Here, ``${pc}$'' subscript refers to the pseudocubic unit cell, while ``$o$'' is used to refer to the orthorhombic unit cell. For (110)$_o$-oriented \dso, the out-of-plane [001]$_{pc}$ direction is equivalent to [110]$_o$, while the in-plane directions [100]$_{pc}$ and [010]$_{pc}$ are equivalent to [$\overline{1}$10]$_o$ and [001]$_o$ respectively (see Supplementary Materials, Figure~\ref{fig:SI_OrthorhombicOrientation}).

\begin{figure}[!htb]
\includegraphics[width=1.0\linewidth]{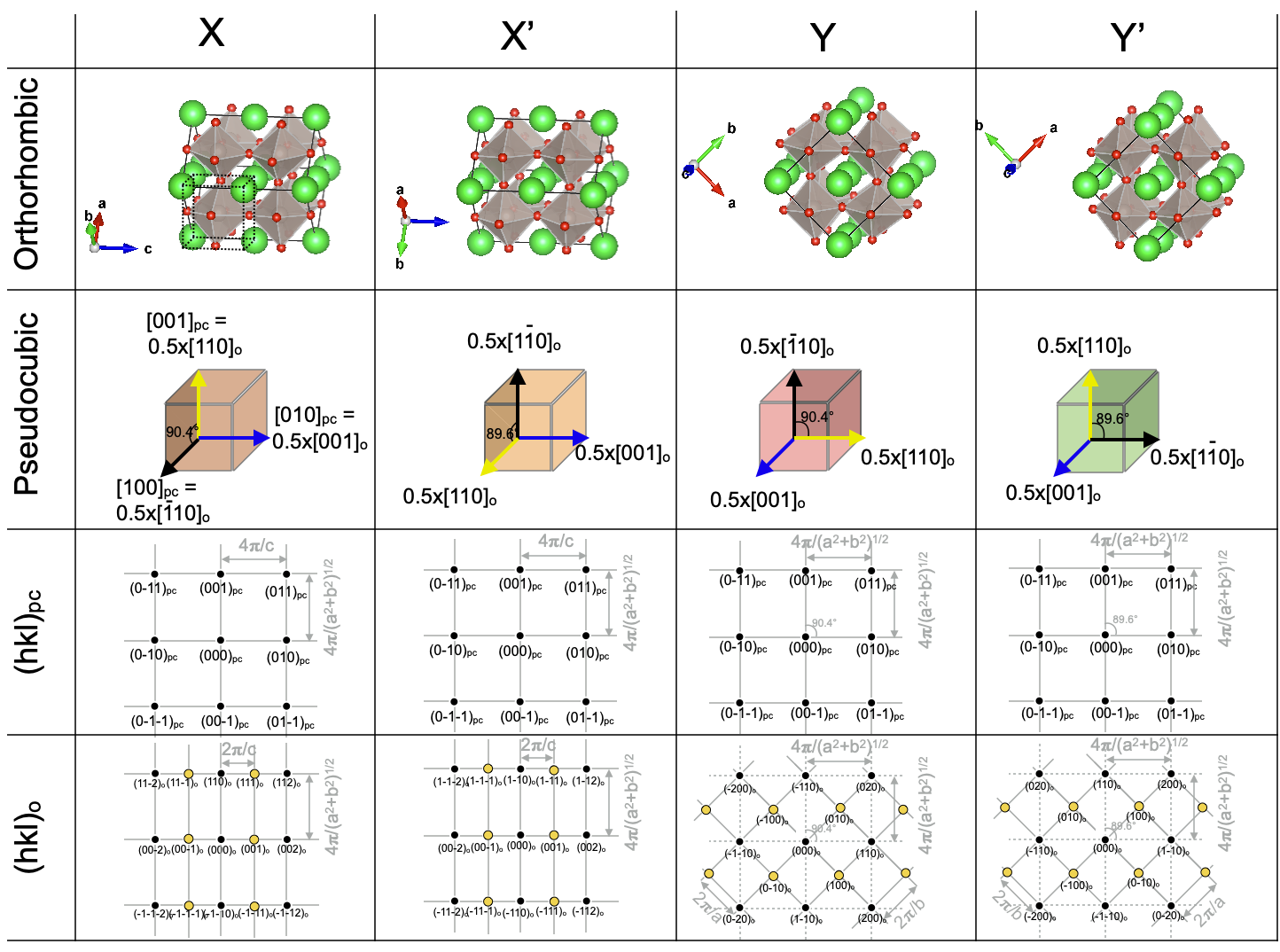}
\caption{\label{fig:SRO_Orientations} Detailed representation of four of the possible orthorhombic orientations of the \sro\ on (110)$_o$-oriented \dso\ substrate (in $Pbnm$ space group). Each column corresponds to a different orientation $X$, $X'$, $Y$ and $Y'$. First row: perspective view, with arrows $\vec{a}\parallel [100]_o$, $\vec{b}\parallel [010]_o$ and $\vec{c}\parallel [001]_o$ corresponding to the axis of the orthorhombic unit cell. Second row: corresponding pseudocubic representation. Third row: reciprocal space representation and (hkl) indices corresponding to the pseudocubic structure. Fourth row: reciprocal space representation and (hkl) indices corresponding to the orthorhombic structure, highlighting the position of the ``half-order'' peaks.}
\end{figure}

The bottom and top electrode, \sro , is orthorhombic with bulk room temperature lattice parameters (in $Pbnm$ space group) $a_o$ = 5.57 \AA , $b_o$ = 5.53 \AA\ and $c_o$ = 7.85 \AA ~\cite{Randall-JACS-1959}, corresponding to the pseudocubic unit cell parameters $a_{pc}$ = $c_{pc}$ = 3.924 \AA , $b_{pc}$ = 3.925 \AA , $\alpha_{pc}$=$\gamma_{pc}$=90$^\circ$, $\beta_{pc}$ = 90.413$^\circ$. According to the Glazer notation, octahedral tilting in orthorhombic \sro\ is described by $a^-a^-c^+$, implying that RuO$_6$ octahedra are rotated in opposite directions by equivalent magnitude along $[100]_{pc}$ and $[010]_{pc}$ (out-of-phase) and in the same direction about $[001]_{pc}$ (in-phase)~\cite{Glazer1972,Woodward-ActaCrystallographicaB-1997}. On (110)$_o$-oriented \dso\ substrate, \sro\ can grow with different possible orientations~\cite{Jiang-MSEB-1998,Wang-NPJQuantMats-2020}, as described in Figure~\ref{fig:SRO_Orientations}, and Supplementary Materials Figure~\ref{fig:SI_OrthorhombicOrientation} and \ref{fig:SI_OrthorhombicOrientation_SRO}.

Last, the \pto\ is ferroelectric below a bulk critical temperature of 765K with a tetragonal structure, and lattice parameters $a$ = $b$ = 3.904 \AA\ and $c$ = 4.152 \AA\ at room temperature. The in-plane strain imposed by \dso\ on \pto\ films can thus be calculated as  $\frac{a_{pc}-a_0}{a_{pc}}=-0.25\%$ along $a_{pc}$ and $\frac{b_{pc}-a_0}{b_{pc}} = -0.16\%$ along $b_{pc}$, where $a_0$ is the extrapolated lattice parameter of \pto\ in the room-temperature cubic paraelectric phase, $a_0$ = 3.957 \AA\ for \pto . To accommodate this strain\footnote{Although the strain with respect to $a_0$ is compressive, the system is usually referred to as being under tensile strain because the lattice parameter of \dso\ is larger than the bulk $a$=$b$ axes of \pto.}, \pto\ thin films on \dso\ at room temperature are expected to be in the $a/c$-phase,  with regions where the $c$-axis points out-of-plane ($c$-domains) as well as regions where it points in-plane ($a$-domains), giving rise to a ferroelastic $a/c$-domain configuration with 90$^\circ$ domain walls. The latter are parallel to the $\{101\}_{pc}$ crystallographic planes, and thus are inclined at about 45$^\circ$ with respect to the film/substrate interface, as predicted in Ref.~\cite{Koukhar-PRB-2001} and demonstrated experimentally (see for example Ref.~\cite{Catalan-NatMat-2011,Nesterov-APL-2013,Highland-APL-2014}). Additionally to these ferroelastic domains, the electrostatic boundary conditions and depolarisation field arising from an incomplete screening of the surface bound charges can lead the $c$-domains to alternate between ``up'' ($c^+$) and ``down'' ($c^-$) orientations. Although the surface bound charges of our \pto\ films are screened by the top and bottom \sro\ electrodes, this screening is incomplete~\cite{Junquera-NAT-2003,Aguado-Puente-PRL-2008,Stengel-NatMat-2009,Li-APL-2017,Hadjimichael-PRM-2020} and the depolarisation field still plays a role. Such a combination of mechanical and electrostatic constraints can then result in flux-closure structures, as observed in strained \pto\ thin films~\cite{Tang-Science-2015,Li-APL-2017,Li-ActaMat-2019,Lichtensteiger-APLMaterials-2023}.

\subsection{Topographic modulation observed at the heterostructure surface using atomic force microscopy}

\begin{figure}[!htb]
\includegraphics[width=0.5\linewidth]{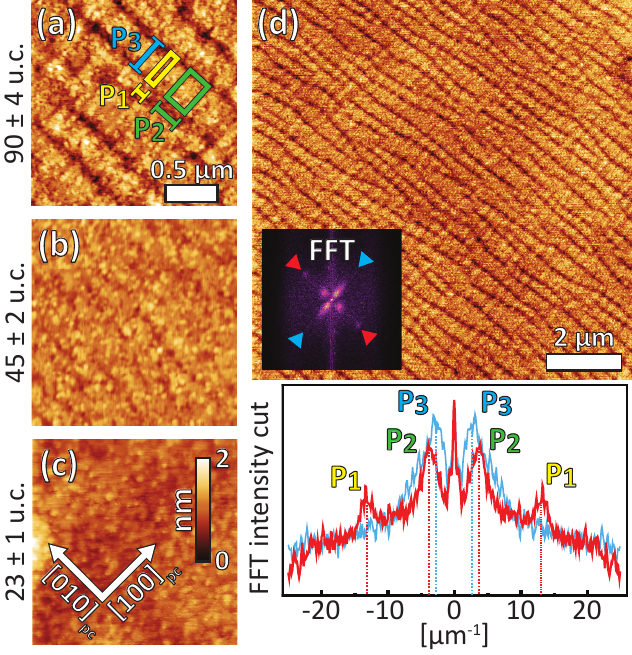}

\caption{\label{fig:AFM} AFM topography images obtained on the different samples. Color scale varies between 0 and 2 nm. The sample orientation was fixed with respect to the substrate pseudocubic axis [100]$_{pc}$//\dso \aOrtho\ and [010]$_{pc}$//\dso \bOrtho. (a-c) 2x2 $\mu$m$^2$ scans for the three samples: (a) 90$\pm$4 u.c., (b) 45$\pm$2 u.c. and (b) 23$\pm$1 u.c. thick \pto\ between top and bottom \sro\ electrodes (55$\pm$2 u.c. thick) on \dso\ substrates, showing that as the \pto\ layer thickness increases, trenches develop at the surface of the \sro\ top layer. (d) A larger 10x10 $\mu$m$^2$ scan for the 90 u.c. thick \pto\ sample displays the anisotropic pattern, with long and deep trenches parallel to the [010]$_{pc}$ axis, and smaller trenches parallel to the [100]$_{pc}$ axis. In the image obtained from the fast Fourier transform of the topography measurement (inset), periodic peaks along [100]$_{pc}$ and [010]$_{pc}$ are visible (see cuts), allowing us to determine the periods. Along [010]$_{pc}$ (red), two periods are visible, P$_1$ = 77 $\pm$ 1 nm and P$_2$ = 280 $\pm$ 3 nm, while along [100]$_{pc}$ (blue), a unique period P$_3$ = 335 $\pm$ 4 nm is visible. These sizes have been drawn on the topography image (a) of the corresponding sample as yellow and green rectangles.}
\end{figure}

Figure~\ref{fig:AFM}(a) shows atomic force microscopy (AFM) images for the three \sro /\pto /\-\sro\ heterostructures grown on \dso . The AFM topography images reveal that as the \pto\ layer thickness increases, trenches develop at the surface of the \sro\ top layer in an organised pattern. For the samples with 23 and 45 u.c. thick \pto\ layers, this pattern is hardly visible and the top \sro\ is smooth. The pattern gets more pronounced and anisotropic with increasing \pto\ layer thickness, with long and deep trenches parallel to the \dso \bOrtho\ axis, and smaller trenches parallel to the \dso \aOrtho\ axis, while the surface roughness stays reasonably low (root mean square (RMS) roughness values ranging from 157 to 393 pm over surfaces of 10 $\mu$m $\times$ 10 $\mu$m).

The pattern that we observe at the surface of the \sro\ top layer is comparable to what has been seen in \pto\ layers grown on \dso\ substrates in Ref~\cite{Nesterov-APL-2013}, attributed to the presence of periodic ferroelastic $a/c$ domains. To extract the period of the distortions visible on the surface of the samples, we calculate the fast Fourier transform (FFT) of the autocorrelation image, as shown in Figure~\ref{fig:AFM}(d) for the sample with the 90 u.c. thick \pto\ layer. Along \dso \bOrtho\ (red), two periods are visible, P$_1$ = 77 $\pm$ 1 nm and P$_2$ = 280 $\pm$ 3 nm, while along \dso \aOrtho\ (blue), a unique period P$_3$ = 335 $\pm$ 4 nm is visible. These periods have been illustrated on the topography image of the corresponding sample as yellow  (with dimensions P$_1$ $\times$ P$_3$) and green (with dimensions P$_2$ $\times$ P$_3$) rectangles. All these values are reported in Table~\ref{table:Periods} (Supplementary Materials).

\subsection{Domain structures observed by scanning transmission electron microscopy}

To better understand the origin of this pattern visible at the surface of the \sro\ top layer, we turned to cross-sectional scanning transmission electron microscopy (STEM) images. The three samples were cut and prepared for STEM measurements to obtain slices in the plane defined by the  \bOrtho\ (horizontal direction) and \cOrtho\ (vertical direction) axes of \dso . STEM images were obtained using bright field (BF), annular bright field (ABF), medium angle annular dark field (MAADF)  and high-angle annular dark field (HAADF) detectors along the \dso \aOrtho\ zone-axis (see section~\ref{section:STEM} for more technical details). 

The domain walls in the \pto\ layers can be directly seen in the STEM images, as shown in Figure~\ref{fig:GPA} (a-c - HAADF images) and in Figure~\ref{fig:sketches} (a-c - BF images), while the \sro\ layers appear rather homogeneous. The \pto\ layers in the three samples studied have different domain configurations, where the expected $a/c$ pattern for the thicker \pto\ layer transforms into a flux-closure pattern for the thinner \pto\ layers (see Ref.~\cite{Lichtensteiger-APLMaterials-2023} for a complete x-ray diffraction based investigation of ferroelectric domain configuration in an extended series of samples).

\subsubsection{Geometric Phase Analysis}

\begin{figure}[!htb]
\includegraphics[width=0.9\linewidth]{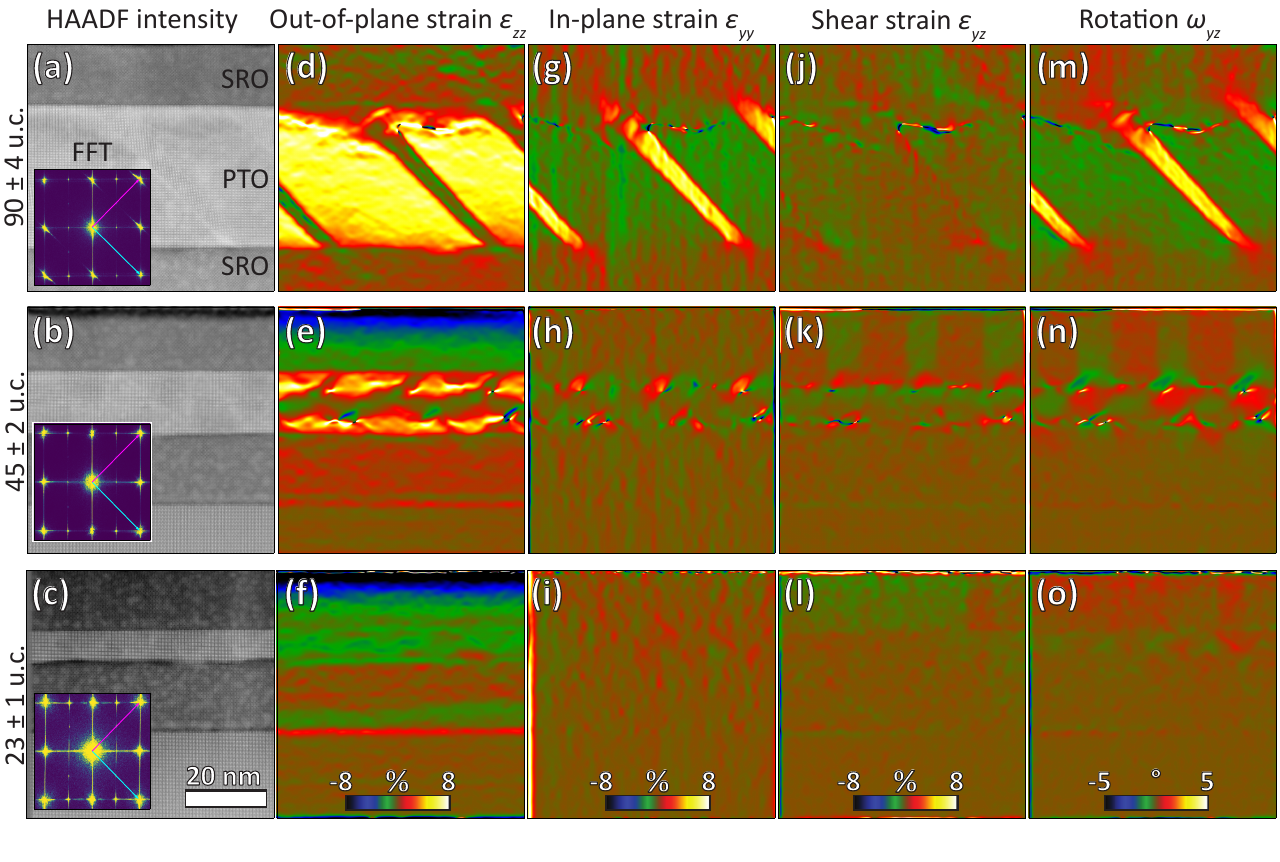}
\caption{\label{fig:GPA} HAADF images and strain maps for the three samples: (top row) 90$\pm$4 u.c., (center row) 45$\pm$2 u.c., and (bottom row) 23$\pm$1 u.c. thick \pto\ between top and bottom \sro\ electrodes (55$\pm$2 u.c. thick) on \dso\ substrates. (a-c) HAADF images, with the inset showing the FFT and the two vectors used for the GPA analysis indicated by the blue and purple arrows. (d-f) Out-of-plane strain $\varepsilon_{zz}$. (g-i) In-plane strain $\varepsilon_{yy}$. (j-l) Shear strain $\varepsilon_{yz}$. (m-o) Rotation $\omega_{yz}$. The strain and rotation values are calculated with respect to a reference lattice here chosen as the substrate. The same intensity scales from -8\% to 8\% (strain) and from -5$^\circ$ to 5$^\circ$ (rotation) are used for the three samples to allow for comparison. Different contrasts are observed in the \pto\ layers, corresponding to the different ferroelastic domain configurations, from $a/c$ for the thickest to flux-closure for the thinner one. The images interestingly reveal an additional contrast appearing only in the top \sro\ layers, for the samples with the 45 u.c. and the 90 u.c. thick \pto\ layers, and propagating all the way to the top surface.} 
\end{figure}

To study the local strain induced by these different domain configurations, we turn to Geometric Phase Analysis (GPA)~\cite{Hytch-Ultramicroscopy-1998}. This is done by taking the FFT of the HAADF-STEM images in Figure~\ref{fig:GPA}, selecting two peaks (here $(01\overline{1})_{pc}$ and $(011)_{pc}$) corresponding to two reciprocal lattice vectors defining the lattice, and getting the inverse Fourier transform containing information about local displacements of the atomic planes along these two vectors. The local strain components are calculated from the derivative of the obtained displacement field: in-plane strain $\varepsilon_{yy}$ (along \dso\bOrtho), out-of-plane strain $\varepsilon_{zz}$ (along \dso\cOrtho), shear strain $\varepsilon_{yz}$ and rotation $\omega_{yz}$.

%PTO
Looking first at the results within the \pto\ layers, we see from the HAADF images and from the GPA maps that the domain patterns vary with \pto\ thickness. For the 90 u.c. thick \pto\ in Figure~\ref{fig:GPA} (top row), we see large regions with a high out-of-plane strain, but low in-plane strain, shear and rotation, corresponding to $c$-domains, i.e. regions where the polarisation is out-of-plane. These regions are separated by narrower features, with high in-plane strain and rotation, but low out-of-plane strain and shear, corresponding to $a$-domains, i.e. regions where the polarisation is in-plane. These results confirm a typical well developed $a/c$-phase. For the 45 u.c. thick \pto\ layer in Figure~\ref{fig:GPA} (center row), the strain map is more complex, with alternating regions with large out-of-plane strain or large in-plane strain close to each interface, and reduced strain at the center of the \pto\ layer, clearly different from an $a/c$-phase (see Supplementary Materials Figure~\ref{fig:SI_HR_45uc}). This pattern is more comparable to the flux closure configuration observed for \pto\ with similar thickness grown without electrodes~\cite{Tang-Science-2015}. Finally, for the 23 u.c. thick \pto\ layer in Figure~\ref{fig:GPA} (bottom row), the \pto\ strain maps are more homogeneous compared to the results obtained for the two other samples. This indicates that for this sample, the distortions related to the ferroelectric/ferroelastic domain configuration in the \pto\ layer are small with respect to the homogeneous strain induced by the substrate. The most pronounced contrast is visible in the rotation map and corresponds to a pattern with a period of $\sim$ 16 nm, in agreement with the value found by XRD (see Supplementary Materials Figure~\ref{fig:SI_RSM} and Table~\ref{table:Periods} for comparison).

%SRO
Concentrating now on the \sro\ layers, we see that the bottom ones are predominantly homogeneous in all three samples. However, this is not the case for the top \sro\ layers, where different contrasts appear for the three different \pto\ thicknesses. While the \sro\ layer on top of the 23 u.c. \pto\ looks rather homogeneous, regions with different shear strain and rotation values alternate in the \sro\ layers grown on top of the 45 u.c. and 90 u.c. \pto\, with boundaries propagating along the [001]$_{pc}$ growth direction. 

For the sample with 90 u.c. \pto , at the interface with each $a$-domain and above the obtuse angle formed by the $a/c$ domain wall, the rotation is positive (red), while it is negative (green) above the acute angle (Figure~\ref{fig:GPA}(m)). The rotation then propagates directly to the top surface along the growth direction \dso\cOrtho . A similar modulation of the rotation is also observed in the \sro\ bottom electrode, with positive rotation below the obtuse angle of the $a/c$ domain wall, and negative rotation below the acute angle. However, for the bottom electrode, this modulation is limited to the vicinity of the interface and does not propagate through the whole \sro\ bottom electrode thickness, most likely due to substrate clamping.

For the sample with 45 u.c. \pto , regions with a positive rotation (red) alternate with regions with a negative rotation (green) (Figure~\ref{fig:GPA}(n)), with a reduced rotation amplitude compared to the sample with 90 u.c. thick \pto , but with sharper boundaries. The period of this pattern follows the period of the ferroelectric domains underneath. 

\subsubsection{Discriminating between $X/X'$ and $Y/Y'$ using fast Fourier transforms}

\begin{figure}[!htb]
\includegraphics[width=0.5\linewidth]{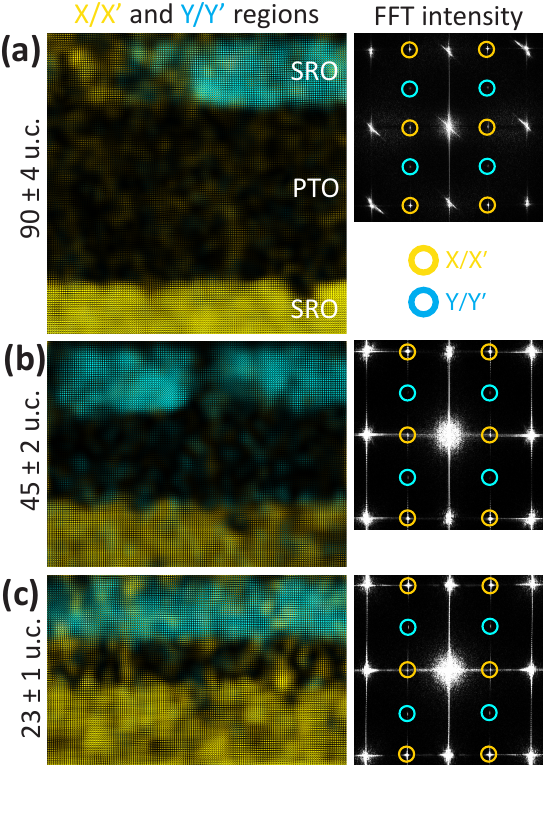}

\caption{\label{fig:FFT_XY} FFT analysis of the \sro\ electrodes orientation in the three samples with (a) 90 u.c., (b) 45 u.c. and (c) 23 u.c. thick \pto\ layers showing the presence of bright peaks corresponding to the pseudo-cubic lattice at \{0 k l\}$_{pc}$ with $k$ and $l$ integer indices, and weaker peaks at half-integer Miller indices \{0 1/2 1\}$_{pc}$, corresponding to $X/X'$ (highlighted in yellow), and \{0 1/2 1/2\}$_{pc}$, corresponding to $Y/Y'$ orientation (highlighted in blue). The color maps are obtained by FFT filtering the orthorhombic superstructures, demonstrating the $X/X'$ orientation for the bottom \sro\ for the three samples, and the $Y/Y'$-orientation for the top \sro\ for the two samples with the thinner \pto\ layers (b,c) and a mixed $X/X'$-$Y/Y'$ for the sample with the thickest \pto\ layer (a).} 
\end{figure}

To better understand the origin of this contrast, one can use the FFT and deduce the orientation of the \sro\ layers from the obtained Bragg peak positions. In the FFT images in Figure~\ref{fig:FFT_XY}, the bright peaks of the pseudo-cubic lattice are clearly visible, corresponding to \{0 k l\}$_{pc}$ with $k$ and $l$ integer indices. Additionally to these peaks, weaker peaks also appear at positions corresponding to half-integer Miller indices \{0 1/2 1/2\}$_{pc}$ highlighted in blue and \{0 1/2 1\}$_{pc}$ highlighted in yellow. These peaks come from the orthorhombic unit cell, which is composed of 4 pseudo-cubic unit cells, as described in Figure~\ref{fig:SRO_Orientations}. The position of the additional peaks is a clear indication of the orientation of the orthorhombic unit cell. The \{0 1/2 1\}$_{pc}$ peaks in the FFT appear when the orthorhombic long axis [001]$_{o}$ is oriented in-plane, parallel to the [010]$_{pc}$ axis, and correspond to the $X$ or $X'$ orientation (note that it is not possible to discriminate between $X$ or $X'$ in this measurement geometry). The \{0 1/2 1/2\}$_{pc}$ peaks on the other hand are the signature of the orthorhombic long axis [001]$_{o}$ being oriented in-plane, parallel to the [100]$_{pc}$ axis, corresponding to $Y$ or $Y'$ orientation. 

By selecting the different half order peaks and reconstructing the images in Figure~\ref{fig:FFT_XY}, we find that the \{0 1/2 1\}$_{pc}$ peaks corresponding to $X/X'$ orientation originate from the substrate and the bottom \sro\ electrode for all three samples, while the \{0 1/2 1/2\}$_{pc}$ peaks corresponding to $Y/Y'$ originate from the top \sro\ electrode. We also note that for the sample with the thickest \pto\ layer, the top \sro\ shows a mixed $X/X'$ and $Y/Y'$ character. Although already highlighting differences in the top \sro\ layers for the different samples, this is not enough yet to explain the contrast observed in the strain and rotation maps in the GPA analysis in the \sro\ top layers for the samples with 45 u.c. and 90 u.c. thick \pto\ layers. This will be further investigated below, where we show that it is possible to discriminate between $Y$ and $Y'$~\footnote{Although the orientation of the lamella does not allow us to discriminate between $X$ and $X'$, we can go one step further by discriminating between $Y$ and $Y'$, the two orientations corresponding to very small horizontal shift in the peak positions.}.

\subsubsection{Discriminating between $Y$ and $Y'$ using fast Fourier transforms}

\begin{figure}[!htb]
\includegraphics[width=0.5\linewidth]{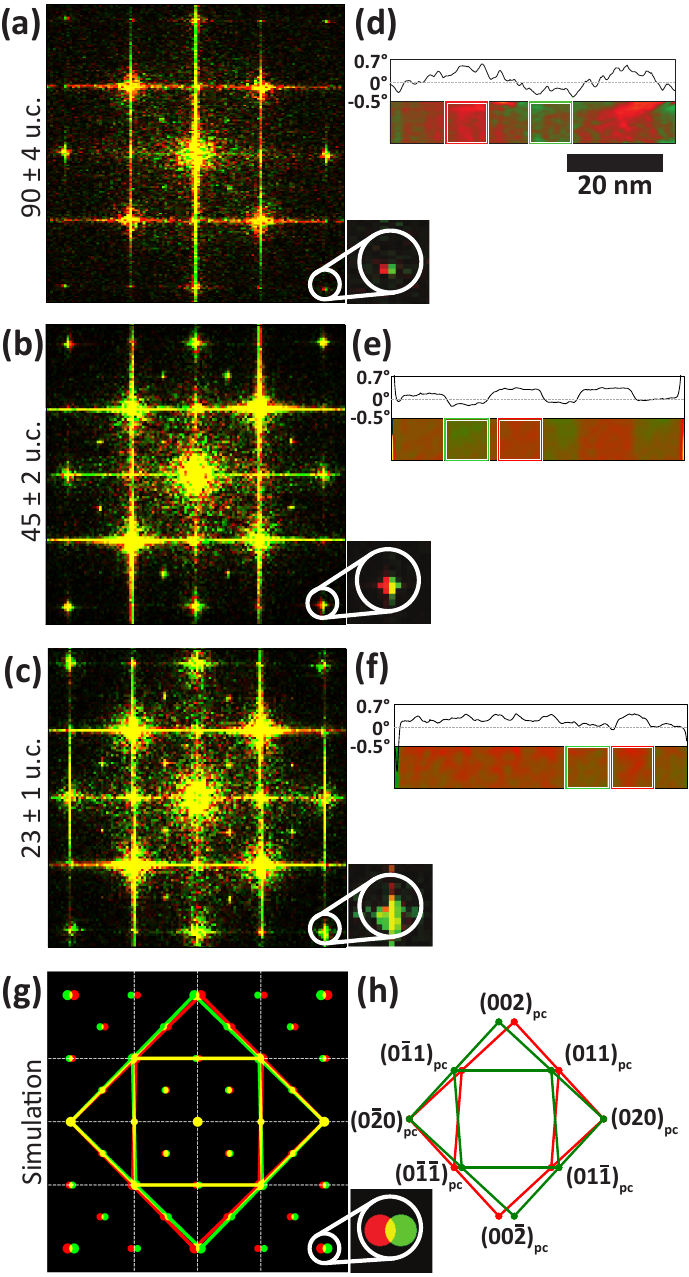}

\caption{\label{fig:FFT_YYprime} FFT of $Y/Y'$ oriented top \sro\ regions of the samples with (a) 90 u.c. \pto\, (b) 45 u.c. \pto\ and (c) 23 u.c. \pto\ (c) compared to the simulation (g) with $a_o=5.62$ and $b_o=5.48$ parameters (in arbitrary units). The colored diffraction patterns of the top \sro\ in (a,b,c) combine the FFT Bragg peaks of the $Y$ domain in red and $Y’$ domain in green, where the regions of interest were selected based on the regions of largest contrast in the GPA rotation maps in (d,e,f).}
\end{figure}

FFT was performed locally on selected regions corresponding to different shear strain and rotation values in the top \sro\ layers of the three samples, as shown in Figure~\ref{fig:FFT_YYprime}. The regions of interest are selected based on the largest contrast in the GPA rotation maps in Figure~\ref{fig:FFT_YYprime}(d-f), and give the colored diffraction patterns in (a-c) for the sample with 90 u.c. (a), 45 u.c. (b) and 23 u.c. (c) \pto\ , respectively, combining the FFT Bragg peaks in red and in green from the two regions. For the samples with 90 u.c. and 45 u.c. \pto\ , the two diffraction patterns do not overlap perfectly, and a very small horizontal shift can be observed (the  $(02\overline{2})_{pc}$ peak is enlarged for clarity). For comparison, the diffraction pattern has been simulated for the $Y$ and $Y'$ orientations by arbitrarily increasing the difference between the $a_o$ and $b_o$ parameters ($a_o=5.62$ and $b_o=5.48$ in arbitrary units), making the difference in the Bragg peak positions more visible: the difference between $Y$ and $Y'$ results in horizontal shifts of peaks, as measured experimentally. No shift is observed for the \sro\ top electrode above 23 u.c. \pto . From this we can conclude that the contrast seen in the GPA strain and rotation maps for the two samples with the thickest \pto\ layers arises from alternating $Y$ and $Y'$ domains. Moreover, Figure~\ref{fig:FFT_YYprime}(d) shows that the transition from $Y$ to $Y'$ domains for the sample with 90 u.c. \pto\ is gradual with a modulation of the rotation following a sinusoidal behavior between 0.3$^\circ$ and -0.3$^\circ$ with a period of approximately 30 nm. In comparison, this transition is much sharper for the sample with the 45 u.c. \pto\ layer, with the modulation of the rotation following a step-like function between 0.15$^\circ$ to -0.2$^\circ$ with a period of approximately 20 nm as seen in Figure~\ref{fig:FFT_YYprime}(e), hinting at the presence of proper twin boundaries between $Y$ and $Y'$ domains. Further discussion and high resolution TEM image of such a twin boundary is shown in Supplementary Materials, Section~\ref{Section:SI_TwinBoundary}.

\section{Discussion}
\label{Section:Discussion_SRO}

\begin{figure}[!htb]
\includegraphics[width=1.0\linewidth]{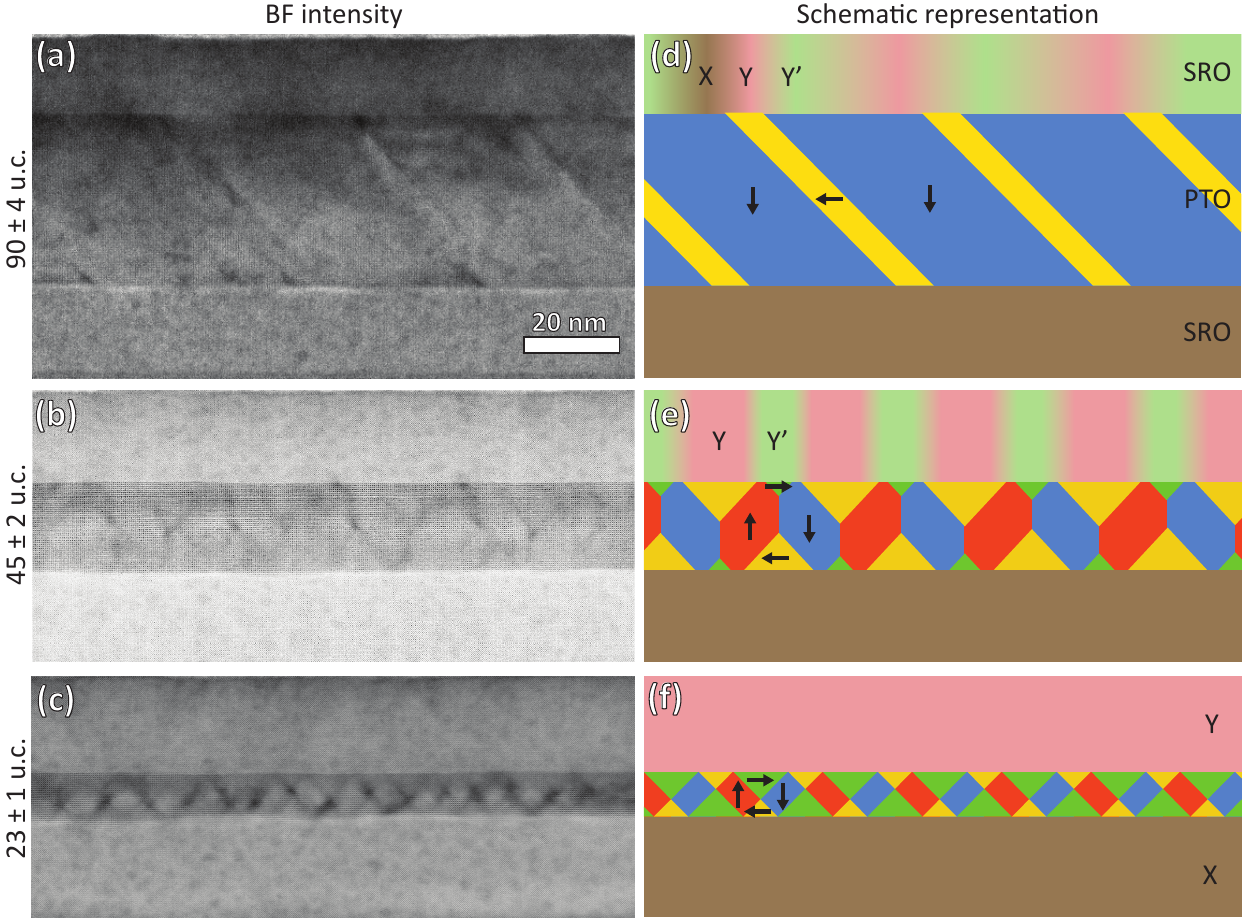}
\caption{\label{fig:sketches} STEM-BF images (a-c) on three different samples, together with the sketches (d-f) representing the different polarisation patterns in the \pto\ layers and the induced crystallographic domains in the top \sro\ layers : (top row) 90$\pm$4 u.c., (center row) 45$\pm$2 u.c. and (bottom row) 23$\pm$1 u.c. thick \pto\ between top and bottom \sro\ electrodes (55$\pm$2 u.c. thick) on \dso\ substrates. From these images, the domain walls are visible in the \pto\ layers, forming the expected $a/c$ pattern for the thicker \pto\ layer, transforming into a flux-closure pattern for the thinner \pto\ layers. In the \pto\ layers, domains with up polarisation are shown in red, down in blue, left in yellow and right in green. In the \sro\ layers, the $X$ (or $X'$) orientation is shown in brown, $Y$ in red and $Y'$ in green.}
\end{figure}

We show that the complex tilt pattern of the \pto\ layer is responsible for the deformation of the \sro\ layer deposited on top, resulting in the periodic pattern visible in the topography by AFM. Our STEM measurements highlight how the domain pattern in the \pto\ layer affects the strain state and crystal orientation in the \sro\ top layer, with clear differences for the three samples with different \pto\ layer thickness.

We find that the top \sro\ layers show a different behaviour for each of the three samples, summarized in Figure~\ref{fig:sketches}, while the bottom \sro\ layers systematically have the same orientation as the \dso\ substrate ($X$-orthorhombic orientation), probably pinned by the interfacial continuity of the oxygen octahedral rotation imposed by the substrate.

For the sample with a 23 u.c. \pto\ layer, where no pattern is visible in the topography, we observe a flux-closure type \pto\ domain structure and correspondingly a null or weakly distorted \sro\ structure, homogeneously in the $Y$-orthorhombic orientation as shown in Figure~\ref{fig:sketches} (bottom row). 

For the sample with a 45 u.c. \pto\ layer, where a pattern is only weakly observed in the topography, we again have a flux-closure type \pto\ domain structure. However, in Figure~\ref{fig:GPA} if we compare the out-of-plane (e) and in-plane (h) strain maps close to the interfaces, the strain pattern locally resembles that of the $a/c$ phase (Figure~\ref{fig:GPA} (g-h)). This points at a nascent $a/c$ phase near the top interface, allowing a more pronounced deformation that translates into a clearly distorted \sro\ structure that shares the same periodicity as the \pto . The top \sro\ thus alternates between $Y$- and $Y'$-orthorhombic orientations - with a period corresponding to the period of the domain pattern in \pto . The mechanism that drives the rotation of the top \sro\ is strain-induced: due to the small shear and rotation in the top layer of \pto , the \pto /top \sro\ interface is not totally flat (left and right inclination). The formation of $Y/Y'$ domains is a good way to minimize the interface strain (Figure~\ref{fig:sketches}(center row)). 

Finally, in the sample with the 90 u.c. thick \pto\ layer, where the topography displays noticeable tilts and trenches, the $a/c$-phase is fully developed, with strong \pto\ lattice heterogeneity that partially imprints on the top \sro\ electrode. Here the top \sro\ exhibits domains with alternating primary $Y$- and $Y'$- orthorhombic orientations, and some $X$ interstitial regions. The positive and negative rotation on the \sro\ correspond to $Y$ and $Y’$ domains as evidenced by the GPA shear strain and rotation maps. In between these $Y$ and $Y’$ domains, a very tiny region with a typical $X$-orientation is often observed in the FFT. This is typically restrained to a 10 nm width domain at the interface between $Y$ and $ Y’$ domains where no rotation was observed on the GPA map. It is different from the case with the 45 u.c. thick \pto\ layer, where sharp interfaces and no $X$ oriented domain were observed between $Y$ and $Y’$ domains. 

Our work demonstrates that the large structural distortions associated with ferroelastic domains propagate through the top \sro\ layer, creating a modulated structure that extends beyond the ferroelectric layer thickness, allowing domain engineering in the top \sro\ electrode. Since there exists a one-to-one correspondence between the structural and magnetic domains~\cite{Wang-NPJQuantMats-2020}, our approach allows magnetic domain engineering in \sro\ thin films through structural domain engineering to be realized. This work paves a new path towards control of magnetic domains via structural coupling to ferroelastic domains.

\section{Experimental techniques}
\label{ExperimentalTechniques}

\subsection{Sample growth}
\label{section:Growth}

The three samples were deposited using our in-house constructed off-axis radio-frequency magnetron sputtering system, equipped with three different guns allowing the deposition of heterostructures and solid-solutions of high crystalline quality.
\pto\ thin films were deposited at 560$^\circ$C, in 180 mTorr of a 20:29 O$_2$/Ar mixture, at a power of 60 W, and using a Pb$_{1.1}$TiO$_3$ target with 10\% excess of Pb to compensate for its volatility.
\sro\ layers were deposited from a stoichiometric target in 100 mTorr of O$_2$/Ar mixture of ratio 3:60, at a power of 80 W. The bottom layer was grown at 640$^\circ$C, while for the top layer the temperature was kept at the growth temperature used for \pto , i.e. 560$^\circ$C, to avoid possible damage of the \pto\ layer.
Huettinger PFG 300 RF power supplies are used in power control mode. The sample holder is grounded during deposition, but the sample surface is left floating. 

\subsection{Atomic force microscopy}

Topography measurements were performed using a {\it Digital Instrument} Nanoscope Multimode DI4 with a {\it Nanonis} controller.

\subsection{Scanning transmission electron microscopy}
\label{section:STEM}

Cross-sectional slices prepared by focus ion beam  allow the imaging of domain structures by scanning transmission electron microscopy. Experiments were acquired on Nion Cs-corrected UltraSTEM200 at 100 kV operating voltage. A convergence angle of 30 mrad was used to allow high-resolution atomic imaging with a typical spatial resolution of 1 \AA. Three imaging detectors in the STEM are used to simultaneously obtain bright field, annular bright field or medium angle annular dark field, and high angle annular dark field images. For ABF-MAADF imaging, the inner-outer angles can be continuously adjusted between 10-20 to 60-120 mrad. Most ABF images were collected with 15-30 mrad and MAADF images with 40-80 mrad angular ranges.  

For the high-resolution HAADF images used to extract GPA, to minimize the influence of the sample drift and environmental noise, a series of fast-scan (low exposure time) HAADF images was taken in the same region; afterward, a script based on Gatan DigitalMicrograph software aligned and summed them together. This technique typically used twenty 4kx4k images with 1$\mu$s exposure time per pixel.

GPA is an algorithm that reconstructs the displacement field $\vec{u}(\vec{r})$ from HAADF images by measuring the displacement of lattice fringes with respect to a reference lattice here chosen as the substrate. GPA thus allows the local strain present in the different layers to be revealed: in-plane strain $\varepsilon_{yy}=\frac{\partial u_y}{\partial y}$ (along [010]$_{pc}$, i.e. perpendicular to the growth direction), out-of-plane strain $\varepsilon_{zz}=\frac{\partial u_z}{\partial z}$ (along [001]$_{pc}$ i.e. along the growth direction), shear strain $\varepsilon_{yz}=\frac{1}{2}\left(\frac{\partial u_z}{\partial y}+\frac{\partial u_y}{\partial z}\right)$ and rotation $\omega_{yz}=\frac{1}{2}\left(\frac{\partial u_z}{\partial y}-\frac{\partial u_y}{\partial z}\right)$. This is particularly useful for the study of the domain configuration in \pto , as the polarisation is related to the strain from the strong strain-polarisation coupling~\cite{Cohen-Ferroelectrics-1992}. At room temperature, \pto\ is tetragonal with the polarisation pointing along the long tetragonal axis. The strain orientation and amplitude therefore indicates the orientation and magnitude of the polarisation. 

We determine the periodicity of the superstructures in the \pto\ and \sro\ layers by measuring the distances between the additional reciprocal space spots obtained after FFT. The accuracy of the measurement was estimated by considering the diffraction spot extension as the lower and upper limit for the superstructure length estimation.

\subsection{X-ray diffraction}

In-house XRD measurements were performed using a {\it Panalytical X'Pert} diffractometer with Cu K$\alpha_1$ radiation (1.5405980 \AA ) equipped with a 2-bounce Ge(220) monochromator and a triple axis detector in our laboratory in Geneva. The $\theta$-2$\theta$ scans were analysed using the {\it InteractiveXRDFit} software~\cite{Lichtensteiger-JApllCryst-2018}. This XRD system is also equipped with a PIXcel 1D detector, used for faster acquisition of the reciprocal space maps. 

\section{Data availability}

The data that support the findings of this study are available at Yareta (DOI).

\section{Acknowlegements}

The authors thank Lukas Korosec and Christian Weymann for support and discussions.

This work was supported by Division II of the Swiss National Science Foundation under projects 200021\_178782 and 200021\_200636. STEM experiments were supported by the EU Horizon research and innovation program under grant agreement ID 823717-ESTEEM3. C.-P. Su acknowledges Taiwan Paris-Saclay doctoral scholarship, which is cost-shared by the Ministry of Education, Taiwan, and the Université Paris-Saclay, France. M.H. acknowledges funding from the SNSF Scientific Exchanges Scheme (Grant Number IZSEZ0\_212990).

\section{Author contributions}

C.L, M.H., A.G. and J.M.T. designed the experiment. C.L., M.H. and L.T. grew the samples and conducted the AFM and XRD measurements and analysis. C.-P.S. and A.G. conducted the STEM measurements. I.G., C.-P.S. and A.G. performed advanced STEM analysis. C.L. wrote the manuscript with contributions from all authors. All authors discussed the experimental results and models, commented on the manuscript, and agreed on its final version.

\section{Bibliography}
%\normalem
\bibliography{biblio}
\bibliographystyle{naturemag}

\newpage
\section*{Supplementary Materials}

\renewcommand{\thepage}{S\arabic{page}} 
\setcounter{page}{1}
\renewcommand{\thesection}{S\arabic{section}}  
\setcounter{section}{0}
\renewcommand{\thetable}{S\arabic{table}} 
\setcounter{table}{0}
\renewcommand{\thefigure}{S\arabic{figure}}
\setcounter{figure}{0}

\section{Orthorhombic orientation}

\begin{figure}[!htb]
\includegraphics[width=0.5\linewidth]{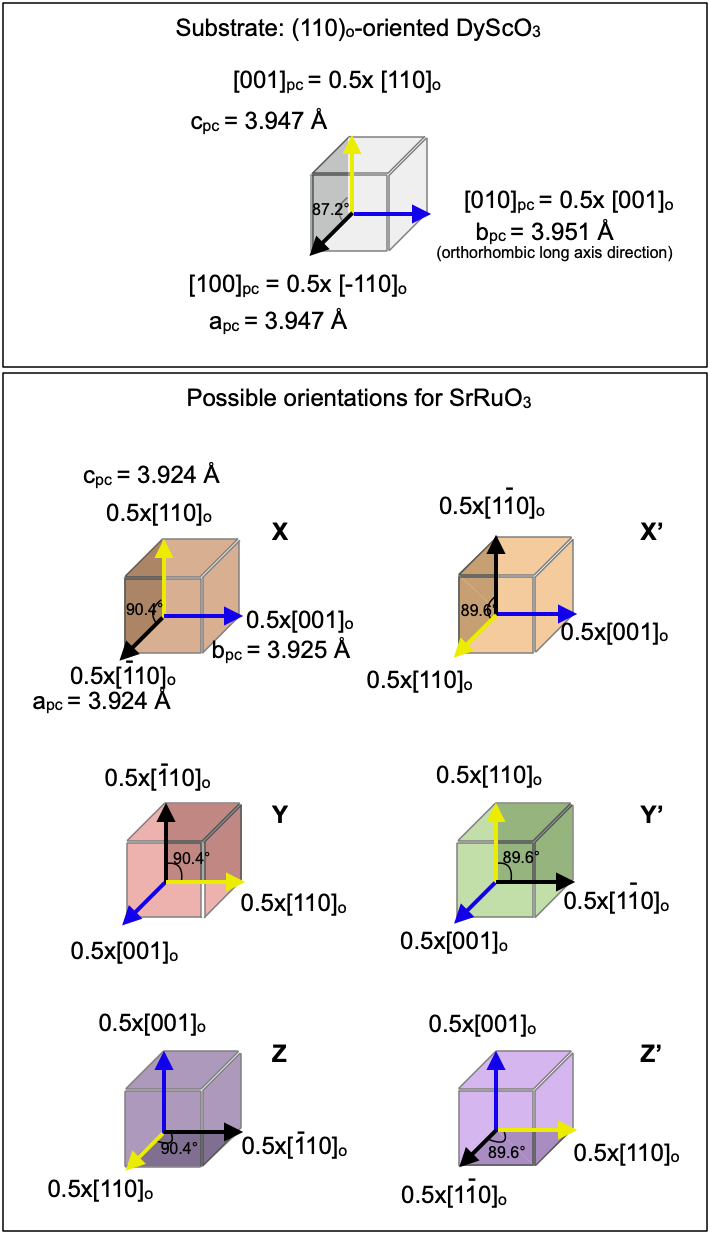}
\caption{\label{fig:SI_OrthorhombicOrientation} Top - Schematic diagram showing the pseudocubic representation of the (110)$_o$-oriented \dso\ substrate and the equivalence between the pseudocubic and orthorhombic axis (in $Pbnm$ notation). Bottom - The six possible orthorhombic orientations of the \sro\ on (110)$_o$-oriented \dso\ substrate (in $Pbnm$ space group). Figure adapted from Ref.~\cite{Jiang-MSEB-1998}. The blue arrow points in the direction of the orthorhombic long axis $c_o$. The yellow and black arrows point along the two diagonals in the ($a_o$,$b_o$) plane, along $[110]_o$ (yellow) and \aOrtho\ (black). The 6 possible orthorhombic orientations are obtained by $\pm \frac{\pi}{2}$ rotations around each of the three pseudo-cubic axes.}
\end{figure}

\begin{figure}[!htb]
\includegraphics[width=1.27\linewidth, angle=90]{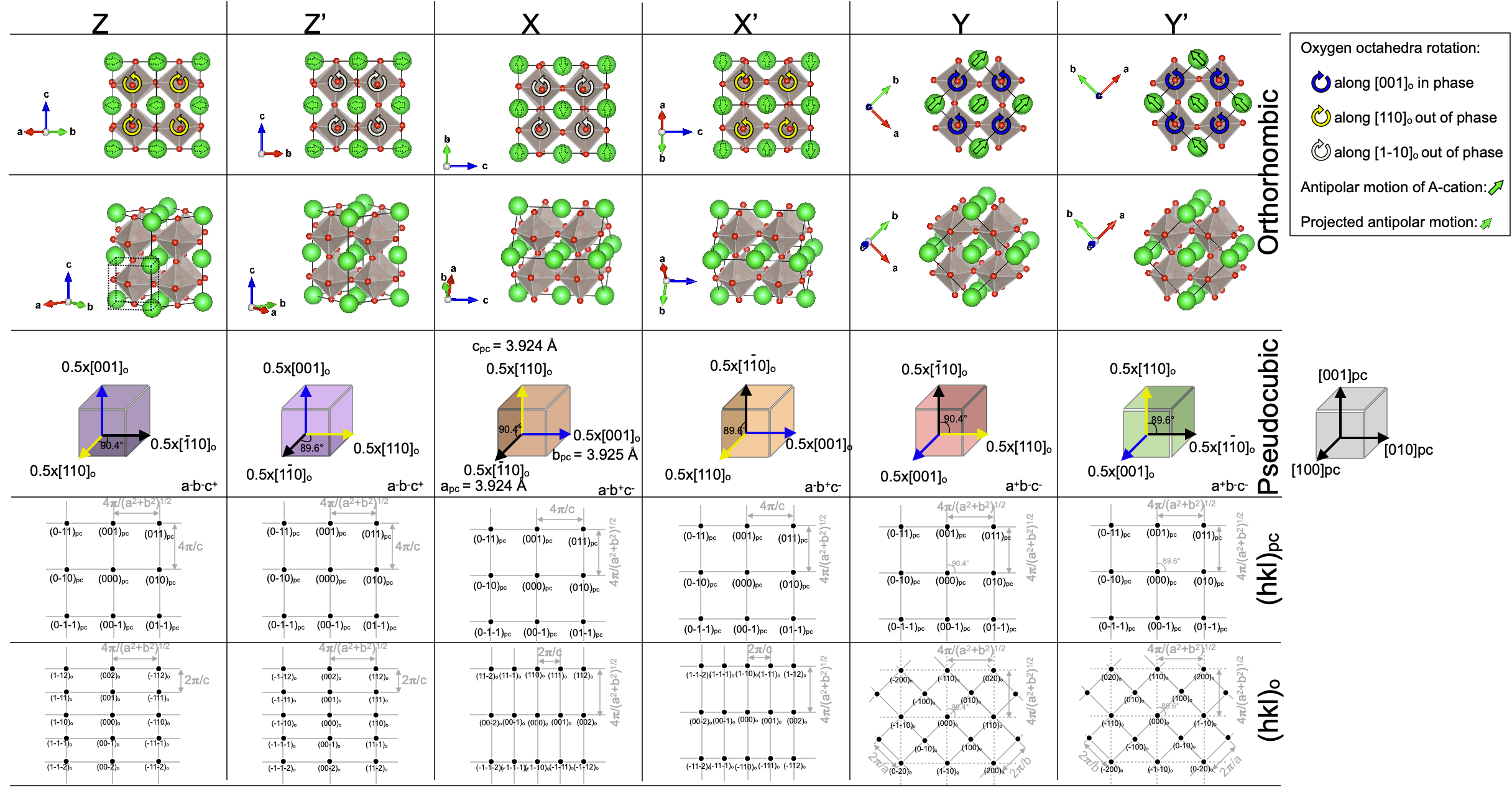}
\caption{\label{fig:SI_OrthorhombicOrientation_SRO} Detailed representation of the six possible orthorhombic orientations of the \sro\ on (110)$_o$-oriented \dso\ substrate (in $Pbnm$ space group).}
\end{figure}

In Figure~\ref{fig:SI_OrthorhombicOrientation_SRO}, the six possible orthorhombic orientations of the \sro\ on (110)$_o$-oriented \dso\ substrate (in $Pbnm$ space group) are represented. Each column corresponds to a different orientation $X$, $X'$, $Y$, $Y'$, $Z$ and $Z'$. First row: projection along [100]$_{pc}$. The oxygen octahedra rotations are represented as blue (in-phase along the [001]$_o$ axis), yellow (out of phase along the [110]$_o$ axis) and white (out of phase along the $[1\overline{1}0]_o$ axis) rotating arrows, and the antipolar motion of the A-cations are represented by green straight arrows. Second row: perspective view. $\vec{a}\parallel [100]_o$, $\vec{b}\parallel [010]_o$ and $\vec{c}\parallel [001]_o$ correspond to the axis of the orthorhombic unit cell. Third row: corresponding pseudocubic representation. Forth row: reciprocal space representation and (hkl)$_{pc}$ indices corresponding to the pseudocubic structure. Fifth row: reciprocal space representation and (hkl)$_o$ indices corresponding to the orthorhombic structure structure, highlighting the position of the ``half-order'' peaks.

\section{X-ray diffraction data for the three samples}

\begin{figure}[!htb]
\includegraphics[width=0.8\linewidth]{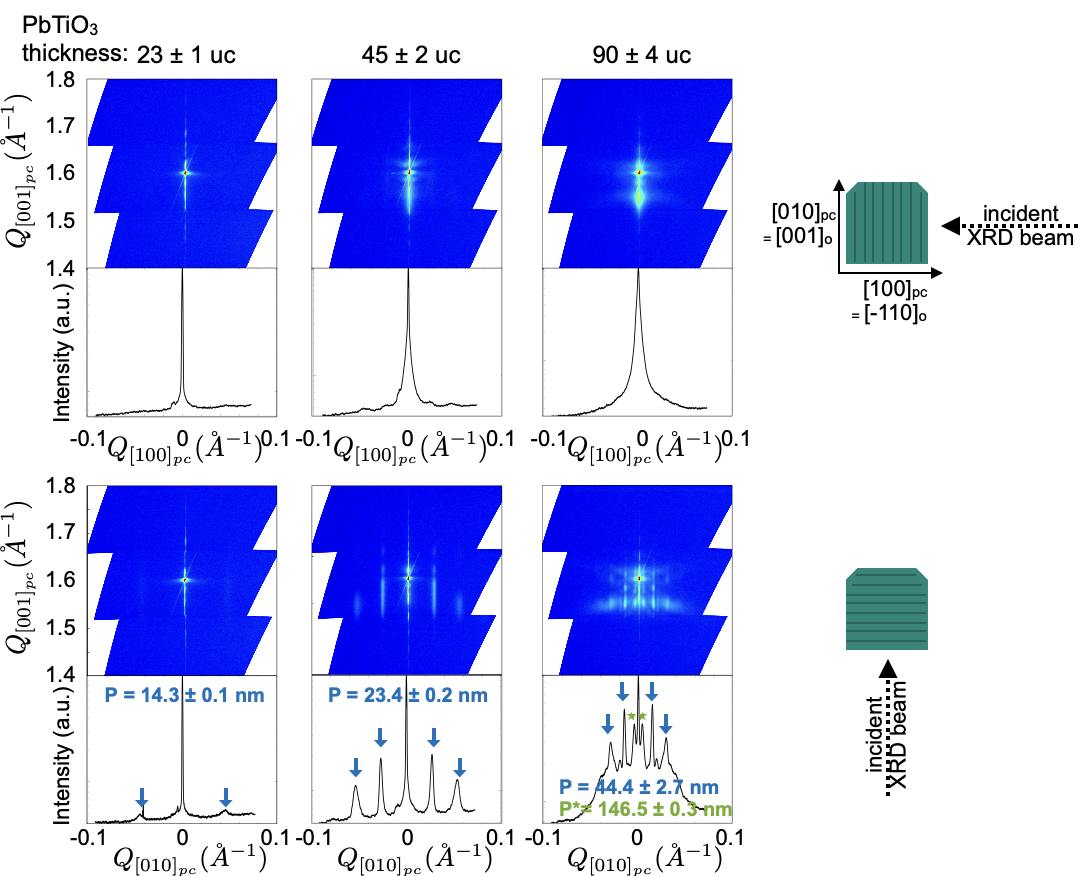}
\caption{\label{fig:SI_RSM} Reciprocal space maps showing the high crystalline quality of the samples studied here - Reciprocal space maps around the (001)$_{pc}$ peak of the substrate in the $Q_{[001]_{pc}}-Q_{[100]_{pc}}$ plane and in the $Q_{[001]_{pc}}-Q_{[010]_{pc}}$ plane. Below each map is the corresponding intensity obtained from a cut at $Q_{[001]_{pc}}$=1.55 \AA$^{-1}$ ($c$-domains), displaying the periodic peaks of the domain pattern. This demonstrates that the periodic pattern is heterogeneous, with periodicity detected only with the incident XRD beam pointing along the \dso\bOrtho -axis. It also shows the different values reported in Table~\ref{table:Periods}, with the additional periodic peak visible for the sample with the thickest \pto\ layer only, marked as a green star.}
\end{figure}

\newpage

\section{Table summarizing the different periods}

\begin{table}[htp]
\begin{center}
\begin{tabular}{|c||c|c|c|c|c|}
\hline
\pto & Period in \pto & Period in \pto & Period in \sro & Period \\
thickness & from XRD & from STEM & from STEM & from topo \\
(u.c.) & (nm) & (nm) & (nm) & (nm) \\
\hline
\hline
23 $\pm$ 1 & 14.3 $\pm$ 0.1 & 13 $\pm$ 1  & n.a.  & n.a.\\
\hline
45 $\pm$ 2 & 23.4 $\pm$ 0.2 & 26 $\pm$ 1 & 25 $\pm$ 1 & n.a.\\
\hline
90 $\pm$ 4 & 44 $\pm$ 3 \ (146.4 $\pm$ 0.3) & 46 $\pm$ 3 & 80 $\pm$ 30 & 77 $\pm$ 1 \ (280 $\pm$ 3)\\
\hline
\end{tabular}
\end{center}
\caption{\label{table:Periods}{\bf Table summarizing the different periods along \dso\bOrtho.}}
\end{table}%

\section{Twin boundary observed by GPA and HAADF in the \sro\ top electrode above the 45 u.c. thick \pto\ layer }

\label{Section:SI_TwinBoundary}

\begin{figure}[!htb]
\includegraphics[width=1.0\linewidth]{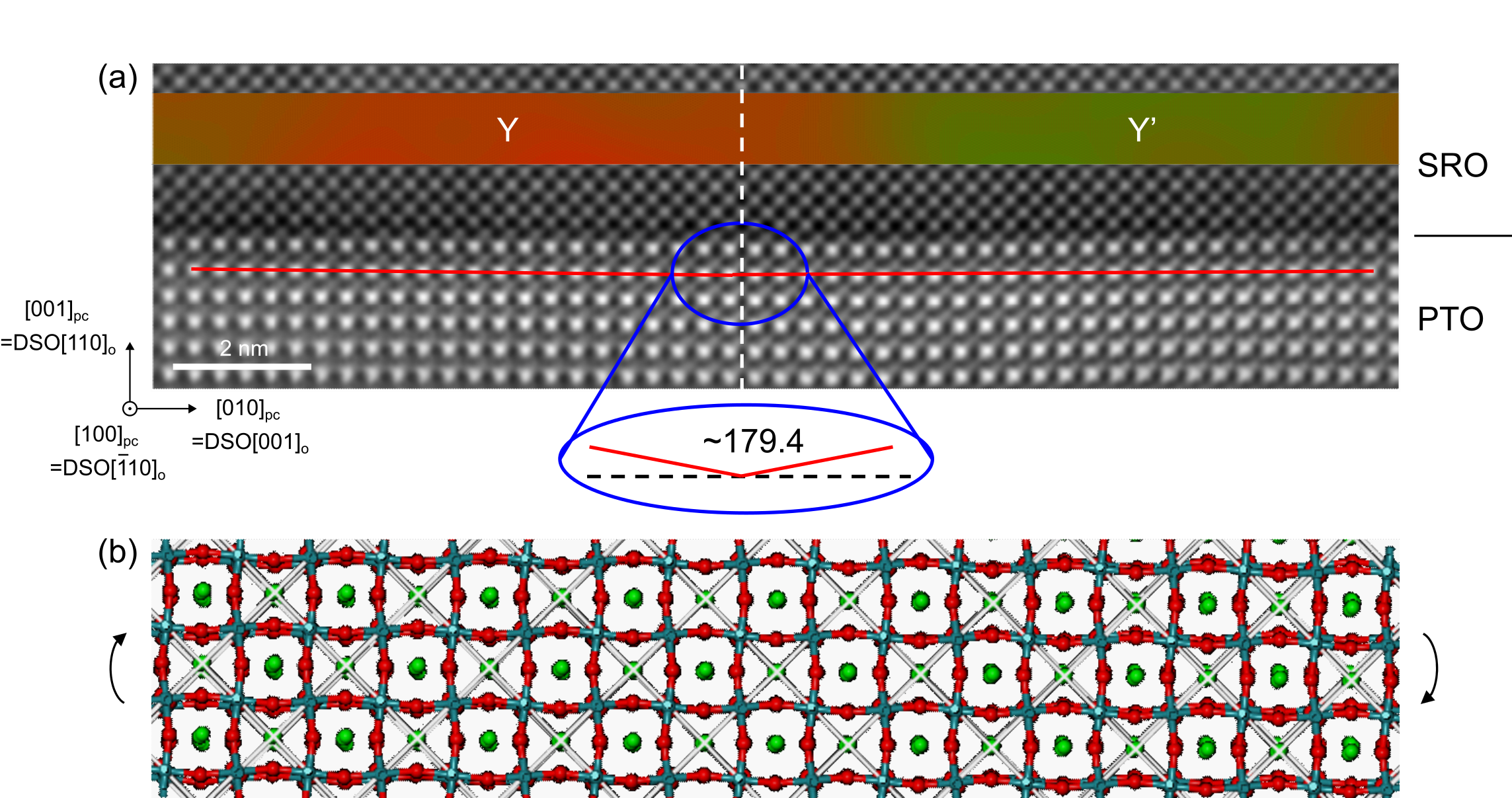}
\caption{\label{fig:SI_HR_45uc} (a) High-resolution HAADF image of the 45 u.c. \pto\ sample. The red lines indicate the atomic layers, showing a small angular difference between the $Y$ and $Y'$ domains. overlaying the corresponding rotation map of GPA. (b) The overlapping of the crystal structure of the $Y$ and $Y’$ domains. It reveals the continuity of cation along [110]$_o$ and a slight clockwise or counterclockwise rotation.}
\end{figure}
 
As mentioned in the paper, the strain map for the 45 u.c. thick \pto\ layer in Figure~\ref{fig:GPA} (center row) is complex, with regions alternating with large out-of-plane strain or large in-plane strain close to each interface, and reduced strain at the center of the \pto\ layer. The succession of sequences close to the interfaces resembles with local $a/c$-domains, but with an offset of about half a period between the sequences at the top and at the bottom interface. In the center of the layer the out-of-plane extension is rather weak, while periodic strain contrasts are present and alternating rotations are visible in the GPA maps. It is clearly different from the expanded $a/c$-phase of the 90 u.c. thick \pto . Horizontal flux-closure quadrants as observed for \pto\ with similar thickness grown without electrodes~\cite{Tang-Science-2015} will give resembling features, but does not explain the local $a/c$-domains at the interface, hinting at a more complicated three-dimensional flux-closure domain. For instance, combination of ``horizontal'' and ``vertical'' flux closures have already been reported in metal-ferroelectric superlattices to minimize the macroscopic polarisation for imperfect electrode while still competing with $a/c$-pattern at longer scale~\cite{Hadjimichael-NatMat-2021}. Such combination can explain the strong flux closure contrast from the HAADF and MAADF STEM images and the presence of nascent local $a/c$-phases at the interface. The complex shear strains and rotations in the \pto\ layer results in a distorted top interface, exhibiting staggered left or right inclination confirmed by the atomic-resolved HAADF image shown in Figure~\ref{fig:SI_HR_45uc}. The interface has a zigzag behaviour due to the presence of the small $a$-domains, inducing \sro\ twinning. This image demonstrates how the growing $a/c$ domains near the top of the \pto\ film imposes the crystallographic domain distribution in the top \sro\ layer. The corresponding rotational and shear patterns of the top \pto\ part are thus corresponding to the $Y$ and $Y'$ domains of the \sro , as shown in Figure~\ref{fig:SI_HR_45uc}, both having the same periodicity (ca. 26 nm), sharing interface connectivity. 

\end{document}